Should alterations in water viscosity be addressed in soil carbon models?

Newton La Scala Jr.[1]; Alexandre Souto Martinez [2,3]; and Kurt Arnold Spokas[4] Daniel Ruiz Potma Gonçalves[5]; Rafael Mazer Etto[6]

1- Universidade Estadual Paulista (UNESP), Faculdade de Ciências Agrárias e Veterinárias / FCAV, Jaboticabal, SP, Brasil

2- Universidade de São Paulo (USP), FFCLRP / Faculdade de Filosofia e Ciências e Letras, Avenida Bandeirantes 3900, 14040-901 Ribeirão Preto, Brasil

3- INCT - Sistemas Complexos, Rua Dr. Xavier Sigaud, 150 - Urca - - CEP: 22290-180, Rio de Janeiro - RJ - Brasil

4- United States Department of Agriculture – Agricultural Research Service – 1991 Upper Buford Circle; St. Paul MN USA 55108

5- Universidade Estadual de Ponta Grossa, Departamento de Fitotecnia e Fitossanidade, Avenida General Carlos Cavalcanti, 4748, 84030-900, Ponta Grossa, PR, Brasil.

6- Universidade Estadual de Ponta Grossa, Departamento de Química, Avenida General Carlos Cavalcanti, 4748, 84030-900, Ponta Grossa, PR, Brasil.

Abstract:

Despite all the efforts, there is no agreement on how temperature affects soil carbon decay and consequently soil $CO_2$ emission, due to overlapping of environmental constraints. To gain further insight into the driving forces of soil microbial processes, we herein examine the abiotic physical environment and its potential influence on microbial activity. In this work we discuss a mechanism which is related to temperature sensitivity of soil carbon stability following a first-order kinetic theory. Soil carbon decomposition is linked to diffusion and consequently to water viscosity, splitting the effects of temperature from viscosity, here we suggest that viscosity could be a controlling factor on bacterial mobility and nutrient diffusion. As a result, viscosity's effect on the potential soil carbon losses is demonstrated and could be an important influence in the feedbacks of climate change on soil carbon cycling kinetics.

While soil carbon comprises only a small fraction of earth's soil (typically less than 10% by mass), it plays a significant role in the global carbon cycle. It is estimated that soils contain approximately 1,500 Pg of carbon, with a huge potential to sequester additional carbon in soil cropland, up to 1.85 PgC per year (Zomer et al., 2017). Nowadays, securing soil carbon stocks (or increasing it) is a priority as soil C is directly linked to several ecological services and has a great potential to aid in climate change mitigation (Zomer et al., 2017; Vermeulen et al., 2019). For instance, in the last COP21 proposed an initiative to increase soil carbon stocks by 0.4% as a mitigation measure for all greenhouse gas emissions due to global anthropogenic sources (Minasny et al., 2017). The strategy should deploy engineering efforts which would deliver increases in soil carbon achieved through communication between science, farmers, and policy makers (Amundson and Biardeau, 2018). However, the spatial distribution of this carbon is not homogeneous across all soils, as there is a concentration of carbon storage, particularly in polar regions (Figure 1). Soil medium has a broad concept (e.g. podzols in equatorial forests and tundra permafrost soils which are located in different ecosystems) and total soil carbon stocks are usually controlled by different factors for diverse biomes (Gonçalves et al. 2021). The spatial distribution is in direct contrast to the above ground biomass distribution (Figure 2a). Typically, the simplified balance of soil carbon is:

$$C = C_{input} - C_{output}. \tag{1}$$

Despite greater terrestrial primary production which should lead to larger soil carbon inputs, in the first layers (until 1 meter depth) soil carbon stocks are notably lower in tropics than in colder regions (Figure 2a). This lack of spatial alignment of carbon inputs with soil carbon storage (Figure 2b), suggests that soil carbon is governed not only by magnitude of the inputs and outputs, but also influenced by the rates of these processes.

Soil carbon output is directly related to microbial activity (aerobic respiration) and soil carbon dynamics is often represented by a simplistic first-order kinetic process (Pingoud and Wagner 2006; Bayer et al., 2006):

$$\frac{dC(t)}{dt} = -kC(t) + I_C, \tag{2}$$

with $C(t)$ representing the soil carbon storage at time $t$, $k$ is the "decay-factor" or decay constant, and $I_C$ is the soil carbon input. Solving for the carbon content at time $t$ results in:

$$C(t) = C_0 e^{-kt} + \frac{I_C}{k}\left(1 - e^{-kt}\right), \tag{3}$$

where $C_0$ is the initial soil carbon content, leading in the steady state to:

$$C_{steady} = \frac{I_C}{k}, \tag{4}$$

with $C_{steady}$ being the steady state of soil carbon stocks.

Therefore, mathematically soils possessing higher decay factors ($k$) result in lower soil carbon stocks at the steady state. Higher $k$ factors are typically seen in tropics when compared to colder regions. Despite the huge changes in carbon inputs around the world, represented in Figure 2a, input amount does not appear to be a dominant driving factor as higher soil carbon stocks are not directly correlated to higher soil inputs (Figure 2b). Therefore, this suggests that the rate of mineralization is the controlling factor and is typically expressed as a function of soil microbial biomass presence, soil temperature, and soil moisture (Leiros et al., 1999; Curtin et al., 2012; Ghimire et al., 2019).

As example, we can examine the Roth-C and Century models, both widely used currently (Coleman, 1999; Parton et al. 2001). The Roth-C model accounts for a compartment Y (Mg C ha$^{-1}$) changes at a rate $(1- e^{-abcdf})$ where a is temperature, b moisture, c is a soil cover constant, d is a material composition constant (e.g., more aliphatic or aromatic) and f is time. Of this decay amount, part is stabilized in soil carbon stocks, being a function of clay content and the rest is considered emitted as $CO_2$. In the Century model, soil respiration is a function of temperature:

$$F_t = (0.17 - 0.68 \times T),$$

where $F_t$ is soil respiration and T is temperature, when carbon is moving from active to slow pools (Leite and Mendonça, 2003). In this case, situations in which soil respiration is not dependent on temperature may not be properly represented by the model framework.

In general, elevated temperatures increase microbial respiration and therefore carbon sequestration tends to decrease with increasing temperature (Qiao et al., 2019). Even though soil microorganisms play a central role in regulating the flow of carbon through soil, information about how abiotic factors interact to drive the carbon use efficiency remains unclear (Domeignoz-Horta et al., 2020).

Temperature changes the water viscosity, and this can affect microbial metabolism and nutrient transport. The viscosity has an important role in maintaining cell function and structure (Pollack, 2001; Dijksterhuis et al., 2007; Sun, 2000). It greatly reduces the physical impacts among biomolecules, increasing its stability and turnover time (Finkelstein et al., 2007; Rauscher et al., 2011). Decreased water viscosity with increasing temperature could result in higher microbial metabolic costs such as production of intracellular solutes (Harris et al., 1981), production of extracellular polysaccharide (Domeignoz-Horta et al., 2020), changes in extracellular enzymatic activity, or alters the stability of low molecular weight biomolecules required for cell functioning and growth (e.g., nicotinamide adenine dinucleotide; NADH) (Cuecas et al., 2016). These effects on the microbial metabolism can influence microbial evolution and growth, impacting the carbon use efficiency and drive changes in microbial diversity (Fierer et al., 2009; Domeignoz-Horta et al., 2020).

Soil microbial biomass varies substantially between environments, in average, bacteria compound ~ 30%, composing up to 10 billion cells per soil gram (Wall, 2012). Recent studies showed great importance of fungi to soil carbon sequestration, stabilizing between 2 – 20% of total carbon stocks, in the form of glomalin bound soil aggregates (Wright et al. 1999; Wang et al. 2017). On the other hand, bacteria transform the soil organic matter through enzyme activity and may have a major influence on carbon respiration rates.

It is estimated that the number of bacterial species per gram of soil varies between $2 \times 10^3$ and $8.3 \times 10^6$ (Gans et al., 2005; Schloss and Handelsman, 2006). Bacteria are the main drivers for soil organic matter decay under aerobic conditions and their physical dispersion influences access to the carbon sources. Bacterial movement through the soil matrix can occur in active (e.g. swimming, growth, swarming and twitching motility) or passive mode (e.g. water-driven transport) (Yang et al., 2018). Although a combination of these two modes of translocation can be used, some bacteria do not have the ability for active movement in the soil solution. In this case, their movement is predominantly controlled by diffusion (Turnbull et al., 2001; Mitchell and Kogure, 2006). However, the direct impacts of water viscosity on bacterial transport remains poorly understood (Griffin and Quail, 1968; Wong and Griffin, 1976; Abu-Ashour et al., 1994; Schneider and Doetsch, 1974; Berg and Turner, 1979; Magariyama and Kudo, 2002; Haines et al., 2009). Mitchell et al. (1991) observed that viscosity accounted for about 26% of the swimming velocity of purple sulfur bacterium (*Chromatium minus*) over a 30°C change in temperature. However, this study was assessing solely the active bacteria movement and not the diffusive transport component.

To gain further insight into the driving forces of soil microbial processes, we herein examine the abiotic physical environment and its potential influence on microbial activity through diffusive transport. Diffusion controlled reaction links the microbial $CO_2$ production rate, consequently the variation in the soil carbon stock in the soil decay ($k$) constant, with the motility of microorganism $k = c\,D$, where $D$ is the diffusion coefficient and $c$ is a proportionality constant (Pilling and Seakins, 1995). The proportionality between $k$ and $D$ describes the connected process of soil carbon decay and $CO_2$ production (Eq. 1-3), which in turn controls carbon soil stock.

The diffusion coefficient is expressed by the Einstein-Stokes relation:

$$D = \frac{K_B T}{\eta(T)}, \qquad (5)$$

with $K_B$ being the Boltzmann constant, $T$ is the soil temperature (in Kelvin) and $\eta(T)$ is the soil water viscosity at temperature $T$. Several factors could affect water viscosity, but for a given soil system temperature would be the main controlling factor. No theoretical formulation exists for the dependency of viscosity and temperature. However, Reid et al. (1987) proposed an empirical expression for the dependency of the viscosity of water as:

$$\eta(T) = 1.856 \times 10^{-11} e^{\left(\frac{4209}{T} + 0.04527 T - 3.376 \times 10^{-5} T^2\right)} mPa\,sec \qquad (6)$$

As seen in Eq. 5, diffusive processes are directly correlated to temperature, with faster rates occurring at higher temperatures. However, the role played by temperature on water viscosity has largely been unaddressed for soil $C$ modeling, despite the greater impact temperature has on altering water's viscosity (Figure 3) and known influence on soil hydraulic properties (e.g., Hopmans and Dane, 1986; Jaynes, 1990) especially on mineral non-hydromorphic and non-frozen soils. For instance, an increase of 11% in soil temperature from 273 K (0 °C, cold regions) to 303 K (30 °C, tropics) is followed by a decrease in 55% in water viscosity from 1.78 to 0.80 mPa.sec. Hence, as seen in Eq. 5, the role of viscosity reduction is much higher (2.2-fold increase due to the reciprocal relationship) than the temperature increase (11%) in changes of diffusion coefficient, D, at elevated temperatures. In soil carbon models, this effect is primarily captured through the $Q_{10}$ relationship on the rate of microbial activity (doubling for every 10 °C change in temperature) (Davidson and Janssens, 2006) and the impact of soil water viscosity changes are not explicated incorporated into soil C models. Furthermore, temperature coefficients (such as $Q_{10}$) has not always adequately describe the temperature dependency of physiological processes particularly involving transport at small spatial scales (Podolsky and Emlet, 1993).

Altering the viscosity of the liquid media has a more significant influence than temperature itself on changes of *D*, the diffusive processes. The plot shown in Figure 3 presents the adjustments of Eq. 5 accounting for changes in soil temperature only, changes of viscosity only (Eq. 6), and combined effect. On one hand, the combined effect of *D* increases the rate of *C* mineralization reactions by microbes with temperature, and thus increasing the output of *C* from the soil system. On the other hand, few have considered that controlling soil water viscosity could provide an engineering strategy to retain soil carbon stocks in tropics and helps to explain results in which crop rotation and fertilization promotes changes in microbial communities (McDaniel and Grandy, 2016; Venter et al. 2016) and increase soil carbon stocks without necessarily higher carbon inputs (Kirkby et al. 2014). Although there have been some attempts at modification of fluid viscosity to reduce water evaporation (Adhikari et al., 2019). There have also been studies documenting increasing water viscosity due to interaction with clay mineral surfaces (Low, 1960) as well as magnetic fields (Ghauri and Ansari, 2006). Soil water viscosity could be a contributing property to reduce rates of diffusion and consequently decay factor (k), leading to higher soil organic matter storage in cold climates.

**Acknowledgement:** Authors acknowledge CNPq (Procs. 309851/2018-1, 304075/2018-3) FAPESP and Capes Agencies (Brazil) for support. This research was also supported by the U.S. Department of Agriculture, Agricultural Research Service. The United States Department of Agriculture is an equal opportunity employer.

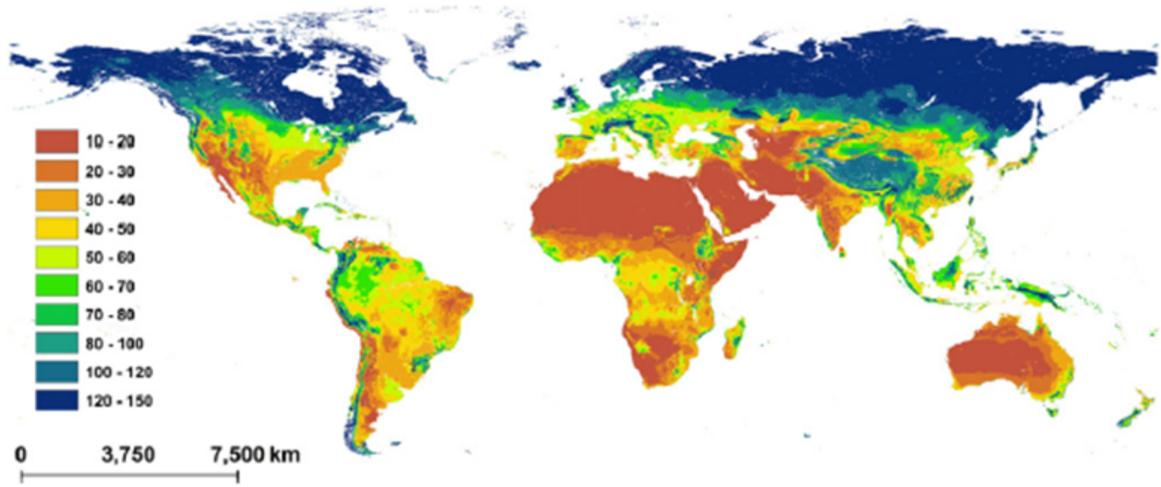

**Figure 1.** World soil C stocks (0-30cm, ton/hectare). From Minasny et al. (2017), based on global datasets of C stock Stockmann et al. (2015).

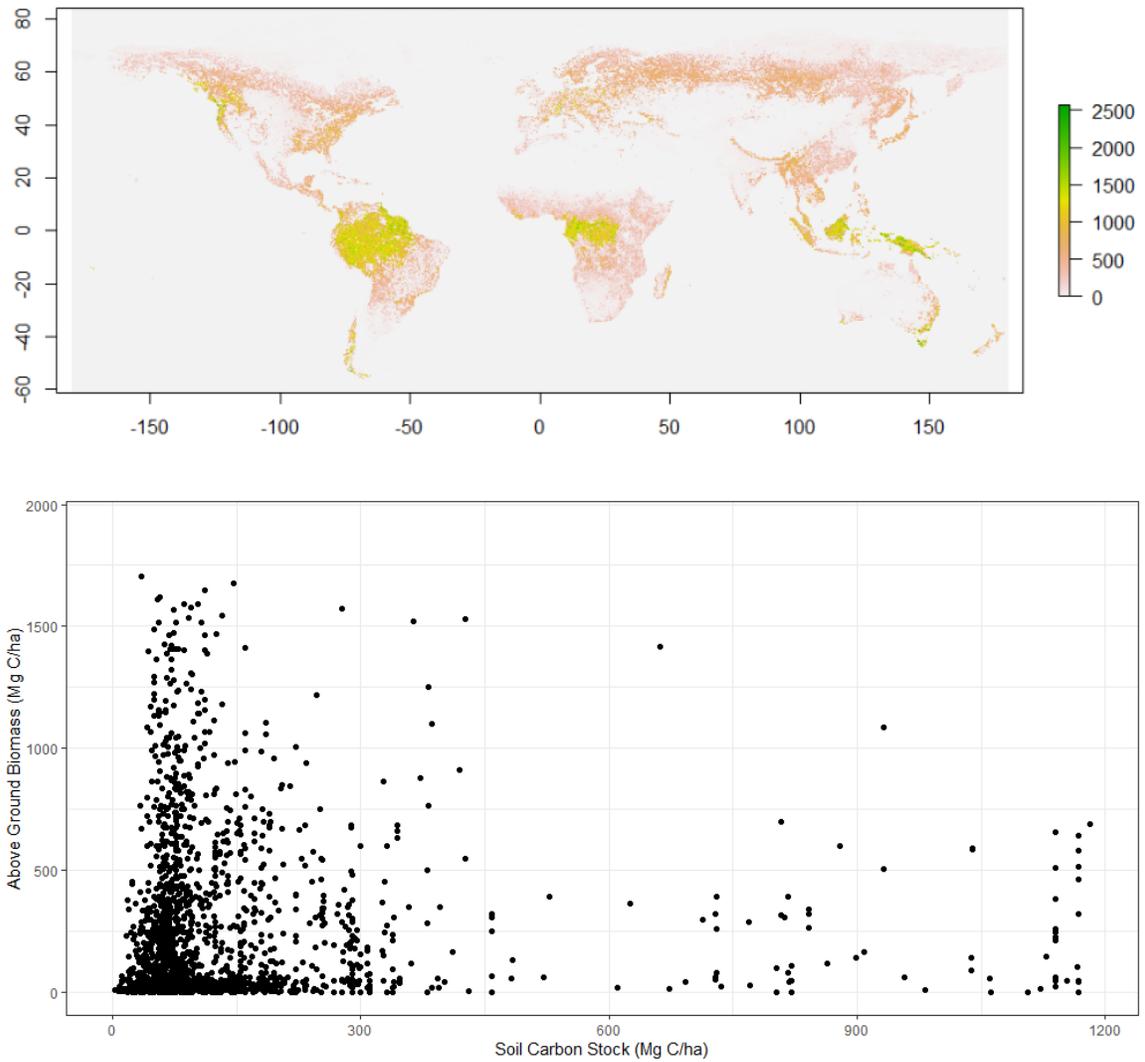

**Figure 2.** a) Global Above ground biomass (Mg ha$^{-1}$) data was taken from Spawn and Gibbs (2020) and b) scatter plot of above ground biomass versus soil carbon stocks which clearly demonstrate the lack of a direct correlation between above round biomass production and current soil carbon stock contents.

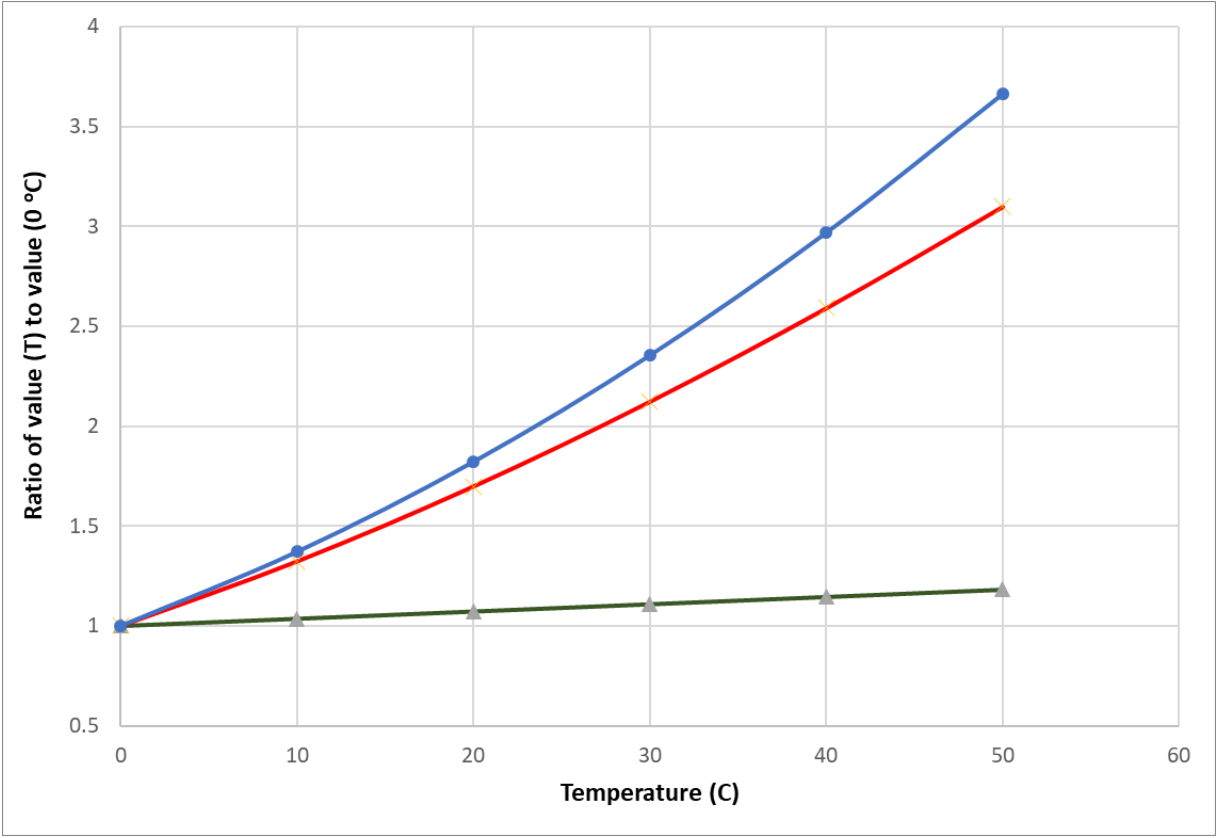

**Figure 3.** Dependency of temperature alone (green), water viscosity alone (red), and their combined effect (blue) on the diffusive processes as a result of alterations in temperature of the system. This impact is expressed as the ratio of the diffusive driving force at temperature T to the diffusive driving force at 0 ºC.